\documentclass[twocolumn,superscriptaddress,showpacs,prd,aps,amsmath,amssymb,nofootinbib]{revtex4-1}
\usepackage{graphicx}
\usepackage{amssymb}
\usepackage{bm}
\usepackage{color}

\newcommand{\Madm}{M_{\rm ADM}}
\newcommand{\MK}{M_{\rm K}}

\newcommand{\Msol}{M_\odot}

\newcommand{\beq}{\begin{equation}} 
\newcommand{\eeq}{\end{equation}} 
\newcommand{\beqn}{\begin{eqnarray}} 
\newcommand{\eeqn}{\end{eqnarray}} 

\newcommand{\na}{\nabla}

\newcommand{\gabu}{g^{\alpha\beta}}

\newcommand{\Tabu}{T^{\alpha\beta}}

\newcommand{\zD}{{\raise1.0ex\hbox{${}^{\ \circ}$}}\!\!\!\!\!D}
\newcommand{\alone}{{\raise0.5ex\hbox{${}^{\ 1}$}}\!\!\!\!\alpha}

\newcommand{\nalam}{\mathrel{\raise0.9ex\hbox{$^\lambda$}\mkern-14mu
\lower0.0ex\hbox{$\nabla$}}}

\newcommand{\zeroD}{{\raise1.0ex\hbox{${}^{\ \circ}$}}\!\!\!\!\!D}

\newcommand{\zLap}{{\raise1.0ex\hbox{${}^{\ \circ}$}}\!\!\!\!\Delta}
\newcommand{\zna}{{\raise1.0ex\hbox{${}^{\ \circ}$}}\!\!\!\!\!\nabla}
\newcommand{\zS}{{\raise1.0ex\hbox{${}^{\ \circ}$}}\!\!\!\!\!S}

\newcommand{\cocal}{{\sc cocal}}

\begin{document}

\title{Modeling differential rotations of compact stars in equilibriums}

\author{K\=oji Ury\=u}
\affiliation{Department of Physics, University of the Ryukyus, Senbaru, Nishihara, 
Okinawa 903-0213, Japan}

\author{Antonios Tsokaros}
\affiliation{Department of Physics, University of Illinois at Urbana-Champaign, Urbana, IL 61801}

\author{Luca Baiotti}
\affiliation{Graduate School of Science, Osaka University, 560-0043 Toyonaka, Japan}

\author{Filippo Galeazzi}
\affiliation{Kurf\"urstenallee 24C, 28211 Bremen, Germany}

\author{Keisuke Taniguchi}
\affiliation{Department of Physics, University of the Ryukyus, Senbaru, Nishihara, 
Okinawa 903-0213, Japan}

\author{Shin'ichirou Yoshida}
\affiliation{Department of Earth Science and Astronomy, Graduate School of Arts and Sciences, 
The University of Tokyo, Komaba, Tokyo 153-8902, Japan}

\date{\today}

\begin{abstract}
Outcomes of numerical relativity simulations of 
massive core collapses or binary neutron star mergers 
with moderate masses suggest formations of rapidly and 
differentially rotating neutron stars.  
Subsequent fall back accretion may also amplify the degree 
of differential rotation.  
We propose new formulations for modeling the differential rotation of those 
compact stars, and present selected solutions of differentially 
rotating, stationary, and axisymmetric compact stars in equilibrium.  
For the cases when rotating stars reach break-up velocities, 
the maximum masses of such rotating models are obtained.  
\end{abstract}

\maketitle

%
%
\section{Introduction}
\label{sec:int}

According to numerical relativity simulations, 
compact stars have a significant amount of differential rotation
in two cases. One is proto neutron stars (PNS) 
formed after the core collapses of supernova progenitors 
around $8-25 \Msol$, and the other is hypermassive neutron 
stars (HMNS) formed after binary neutron star (BNS) mergers \cite{LIGO,Simulation}.  
The rotation curve of PNS, roughly speaking the angular velocity profile 
$\Omega$ as a function of the cylindrical radial coordinate 
$\varpi = r \sin\theta$, may depend on various factors including 
initial spins, magnetic field configurations, and equations of 
state (EOS), and may evolve in time with a timescale 
longer than the dynamical one.  

It has been reported in recent core collapse simulations that 
in the range $\sim 20-30$km from the rotation axis the rotation curves are 
monotonically decreasing of about a few tens of \% from the value 
at the rotation axis \cite{SNrot}.  
%
%
The rotation curves of the HMNS formed after binary neutron star 
mergers are more complex.  Recent simulations \cite{BNSremnant} suggest that 
the profile of the rotation curve $\Omega(\varpi)$ increases of about 
a few tens to a few hundreds of \% of the central value $\Omega_{\rm c}$ and 
then decreases, independently of EOS.  
Not only the above dynamical process, but also the subsequent fall back 
accretion may amplify the degree of differential rotations \cite{Fallback}.
If the matter from a fall back disk accretes onto the equatorial 
surface of a nascent neutron star to spin it up, the angular velocity 
of the outer part of neutron star could become even faster than 
the inner part.  

%
%
\begin{figure*}
\begin{center}
\includegraphics[height=48mm]{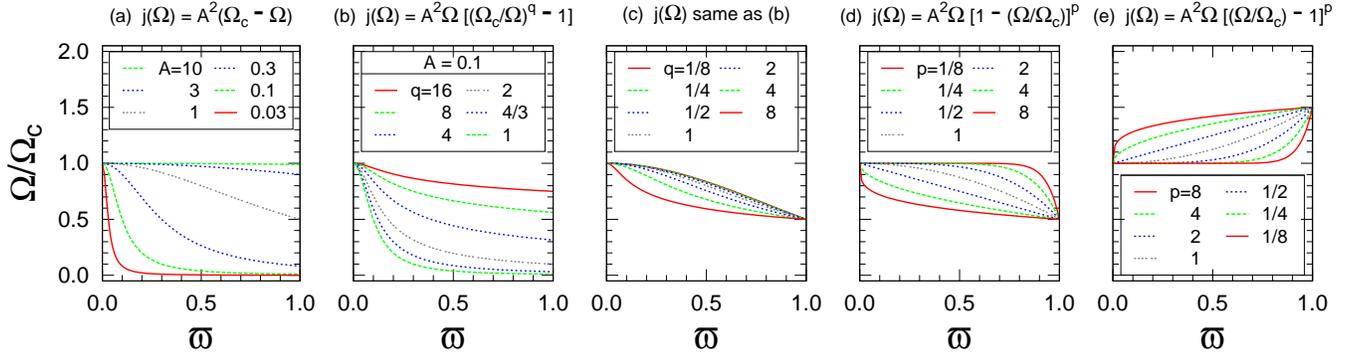}
\caption{
Rotation curves $\Omega(\varpi)$ for the rotation laws 
(\ref{eq:rotlawjc}), (\ref{eq:rotlaw0}) and (\ref{eq:rotlaw12}) 
in the Newtonian limit.  
Panel(a): $j$-constant law (\ref{eq:rotlawjc}) 
(whose $\Omega(\varpi)$ is the left expression
in Eq.(\ref{eq:rotcurve012}) with $q=1$).  
Panel(b): Eq.~(\ref{eq:rotlaw0}) with $A=0.1$ 
($\Omega(\varpi)$ is the left expression in Eq.~(\ref{eq:rotcurve012})).  
Panel(c): same as panel(b), Eq.~(\ref{eq:rotlaw0}) 
but with $\Omega_{\rm eq}/\Omega_{\rm c}=0.5$.  
Panel(d): a new rotation law (\ref{eq:rotlaw12}) with 
a negative sign ($\Omega(\varpi)$ is the right expression in 
Eq.~(\ref{eq:rotcurve012}) with a negative sign).  
Panel(e): a new rotation law (\ref{eq:rotlaw12}) with 
a positive sign ($\Omega(\varpi)$ is the right expression in 
Eq.~(\ref{eq:rotcurve012}) with a positive sign).  
For the cases of panel(c)-(e), the parameter $A$ 
is determined by setting $\Omega_{\rm eq}/\Omega_{\rm c}=0.5$, 
where $\Omega_{\rm c}$ and $\Omega_{\rm eq}$  are the angular velocities 
at the rotation axis and at the equatorial surface, respectively.
In each panel, the order of the labels in the legends (from top left to bottom right) 
correspond the order of the curves in the plot (from top to bottom).
}  
\label{fig:rotcurve012}
\end{center}
\end{figure*}
\begin{figure}[h]
\begin{center}
\includegraphics[height=46mm]{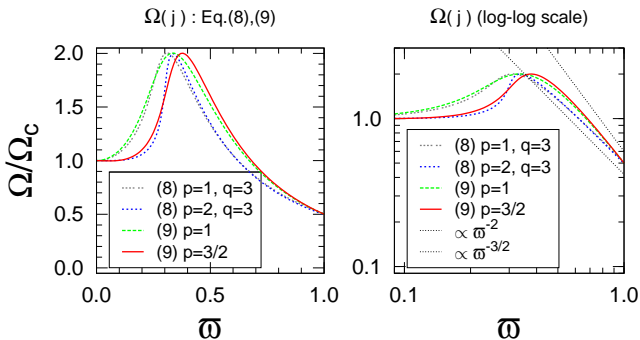}
\caption{
Selected rotation curves $\Omega(\varpi)$ for the rotation laws 
(\ref{eq:rotlawOJ}) and (\ref{eq:rotlawOJjco}) in the Newtonian limit.  
Left and right panels are the same curves in linear and 
log-log scales, respectively.    
The parameters $A$ and $B$ are determined by setting 
$\Omega_{\rm eq}/\Omega_{\rm c}=0.5$ and $\Omega_{\rm max}/\Omega_{\rm c}=2$,
where $\Omega_{\rm max}$ is the maximum angular velocity 
at a given point on the equatorial plane.  
}  
\label{fig:rotcurveOJ}
\end{center}
\end{figure}

Modeling such differential rotations in the relativistic regime 
is important for accurately computing stationary and 
axisymmetric equilibriums of the above mentioned astrophysical 
compact objects.  Such equilibrium solutions will be 
useful for studying their long time evolutions 
in thermal/viscous timescales, for studying their dynamical or 
secular stabilities, and for providing initial data for 
numerical relativity simulations \cite{RNSreview}.
Stationary and axisymmetric models of rapidly rotating 
stars can be calculated most straightforwardly by 
deriving an analytic first integral of the Euler equation, 
and by simultaneously solving the first integral and the 
gravitational field potentials using a certain iterative, 
self-consistent, numerical method \cite{SCF,RNS}.  
In Newtonian gravity, the first integral can be derived 
assuming a one-parameter EOS for 
the thermodynamic variables (namely, a barotropic fluid), 
or the flow field to be $v^a = \Omega \phi^a$, where 
the angular velocity $\Omega$ depends only on $\varpi$, 
$\Omega=\Omega(\varpi)$. Either one of these choices implies 
the other condition \cite{1978trs..book.....T}.\footnote{
In this paper, the Latin indices denote spatial vectors or 
tensors, the Greek indices spacetime ones, and geometric 
units $G=c=1$ with solar mass $\Msol=1$ are used.}  

In the case of relativistic gravity, the relativistic Euler 
equation associated with the timelike and rotational 
Killing fields $t^\alpha$ and $\phi^\alpha$ can be 
analytically integrated for the circular flow, 
$u^\alpha=u^t(t^\alpha + \Omega \phi^\alpha)$, 
where $u^\alpha$ is the 4 velocity of the perfect fluid.  
Different from the Newtonian case, however, 
the integrability condition can not be expressed as 
$\Omega$ to be a function of coordinate $\varpi$, 
but as the relativistic specific angular momentum 
$j:=u^t u_\phi$ ($u_\phi:=u_\alpha\phi^\alpha$) to be 
a function of $\Omega$, $j=j(\Omega)$ \cite{RNSreview}.  
For this condition, it is less clear than the Newtonian 
case how one should set the form of the integrability 
condition $j(\Omega)$.  Because of this, only limited 
types of differential rotations have been investigated 
for relativistic stars 
\cite{RNSdiff,Baumgarte:1999cq,Morrison:2004fp,Kaplan:2013wra}.  

The differential rotation law used in most of previous 
works is the so called $j$-constant law 
\beq
j(\Omega) = A^2(\Omega_{\rm c}-\Omega), 
\label{eq:rotlawjc}
\eeq
where $\Omega_{\rm c}$ and $A$ are constants. 
One of the well known applications of this rotation law is 
the computation of HMNS formed after binary neutron star mergers.  
Baumgarte, Shapiro, and Shibata (BSS) \cite{Baumgarte:1999cq} have 
shown that this rotation law can support stars having nearly 
twice of the maximum mass of the non-rotating star for the same EOS.  
This work was extended by several authors using the same 
rotation law \cite{Morrison:2004fp,Kaplan:2013wra}.  
However, this rotation curve (approximately, the curve labeled $A=1$ 
in panel (a) of Fig.\ref{fig:rotcurve012}) is even 
qualitatively different from the ones resulting from the simulations 
mentioned above \cite{BNSremnant}.  
To our knowledge, differential rotation laws different from 
Eq.~(\ref{eq:rotlawjc}) have been used only in 
\cite{Galeazzi:2011nn,Nick,Uryu:2016dqr} for computing compact stars, 
and in \cite{Mach_Malec} for a rotating self-gravitating disk around 
a point source in first order post-Newtonian gravity.  

In this paper, we introduce new formulations for modeling 
realistic rotation curves of differentially 
rotating compact stars extending our previous 
works \cite{Galeazzi:2011nn,Uryu:2016dqr}
and present the first results of the set of equilibrium 
solutions.  
%

%
%
\section{New formulations for the relativistic differential rotations}
\label{sec:method}

The relativistic Euler equations are derived from 
the transverse components of the conservation laws, 
$\na_\beta \Tabu=0$, 
with respect to the 4 velocity $u^\alpha$, 
where the perfect fluid stress energy tensor is written 
$\Tabu = (\epsilon+p)u^\alpha u^\beta+p\gabu$.  
Here, $\epsilon$ is the energy density, $p$ the pressure, 
and $\gabu$ the spacetime metric.  
Applying the symmetries along the timelike and rotational 
Killing fields $t^\alpha$ and $\phi^\alpha$, they are written 
\beq
\na_\alpha \ln \frac{h}{u^t}
\,+\, u^t u_\phi \na_\alpha \Omega 
\,-\, \frac{T}{h}\na_\alpha s \,=\,0, 
\label{eq:Euler}
\eeq
where $h$ is the relativistic enthalpy defined by $h=(\epsilon+p)/\rho$, 
\footnote{The relativistic enthalpy $h$ 
satisfies the local thermodynamic relation $dh=Tds+dp/\rho$.}
$\rho$ the baryon rest mass density, 
$T$ the temperature, and $s$ the specific entropy.  In this article, 
we assume a homentropic fluid $s={\rm constant}$, although 
an extension to a more general barotropic fluid is straightforward.  
Then, with the integrability condition $j:=u^t u_\phi =j(\Omega)$, 
Eq.~(\ref{eq:Euler}) is analytically integrated as 
\beq
\frac{h}{u^t}\exp\left[\int jd\Omega\right]\,=\,{\cal E}, 
\label{eq:1stint}
\eeq
where ${\cal E}$ is a constant.

The choice of the function form of the integrability condition $j(\Omega)$ 
is the key for modeling various rotation curves in relativistic stars.  
In our previous paper \cite{Uryu:2016dqr}, the following rotation law 
has been introduced 
\beq
j(\Omega;q,A) = A^2 \Omega\left[\left(\frac{\Omega_{\rm c}}{\Omega}\right)^q - 1 \right].  
\label{eq:rotlaw0}
\eeq
In addition to this, we propose new rotation laws, 
\beq
j(\Omega;p,A) = A^2 \Omega\left[\mp\left(\frac{\Omega}{\Omega_{\rm c}} - 1\right) \right]^p.  
\label{eq:rotlaw12}
\eeq
The rotation laws (\ref{eq:rotlaw0}) and (\ref{eq:rotlaw12}) can be combined 
in a general form 
$j(\Omega;p,q,A) = A^2 \Omega\left[\mp\left((\Omega_{\rm c}/\Omega)^q - 1 \right) \right]^p$.  
However, we separate them as (\ref{eq:rotlaw0}) and (\ref{eq:rotlaw12}), 
because they are general enough, and because the integrals of $j(\Omega;p,q,A)$ 
that appear in Eq.~(\ref{eq:1stint}) involve hypergeometric functions 
for arbitrary values of the indices $(p,q)$, which may be inconvenient 
for a numerical code.  
The minus and plus signs of the rotation law (\ref{eq:rotlaw12}) give 
monotonically decreasing and increasing rotation curves, respectively, 
by choosing a set of positive real roots for $\Omega$ which satisfies  
$\mp\left(\frac{\Omega}{\Omega_{\rm c}} - 1\right)>0$ (see below).  
We usually choose the power indices $p$ and $q$ in Eqs.~(\ref{eq:rotlaw0}) 
and (\ref{eq:rotlaw12}) to be positive real numbers.  

In the Newtonian limit, $\Omega(\varpi)$ can be 
solved from $u^t u_\phi = \varpi^2\Omega = j(\Omega)$ for each rotation law 
(\ref{eq:rotlaw0}) and (\ref{eq:rotlaw12}), respectively, as 
\beq
\frac{\Omega}{\Omega_{\rm c}}= \left( 1+\frac{\varpi^2}{A^2}\right)^{-1/q}, 
\ \ \mbox{and}\ \ 
\frac{\Omega}{\Omega_{\rm c}}= 1\mp\left(\frac{\varpi^2}{A^2}\right)^{1/p}.  
\label{eq:rotcurve012}
\eeq
From Eqs.~(\ref{eq:rotcurve012}), it becomes clear 
that the role of the parameter $A$, which has the dimension of length, 
is to set the radius where the rotation curve changes from a constant 
$\Omega_{\rm c}$ to a certain differential rotation at a radius around 
$\varpi \sim A$.  As discussed in \cite{Uryu:2016dqr}, 
for the case of rotation law (\ref{eq:rotlaw0}), which corresponds to 
the first rotation curve in Eq.~(\ref{eq:rotcurve012}), 
$\Omega$ becomes a power of $\varpi$, namely 
$\Omega\sim \varpi^{-2/q}$, in the Newtonian limit for $A \alt \varpi \leq R_0$, 
where $R_0$ is the equatorial radius of the compact star.  
For example, it becomes the Kepler rotation law for $q=4/3$ and the $j$-constant law 
for $q=1$, as shown in panel (b) of Fig.~\ref{fig:rotcurve012}.  
In the regime of strong gravity, rotation profiles are modified 
because of relativistic effects (including the choice of coordinate conditions).  
Some examples are shown for selected solutions in the later section.

This rotation law (\ref{eq:rotlaw0}) may be used in a different manner: 
by adjusting the constant parameter $A$ for a given slope index 
parameter $q$, one can fix the value of $\Omega$ at a given point to 
a given value.  
For example, we can fix the ratio of $\Omega$ at the equatorial surface 
$\Omega_{\rm eq}$ to its central value $\Omega_{\rm c}$, and 
vary the slope of the rotation curve $\Omega(\varpi)$.  
In panel (c) of Fig.\ref{fig:rotcurve012}, such rotation curves 
$\Omega(\varpi)$ of Eq.~(\ref{eq:rotlaw0}) in the Newtonian limit 
(the first equation in (\ref{eq:rotcurve012})) are plotted 
for $\Omega_{\rm eq}/\Omega_{\rm c}=0.5$.  
When the positive index $q$ is decreased, one might expect that 
the rotation curves become more and more convex upward.  However, as 
seen in panel (c), it is not the case: the slopes of the rotation curves 
with fixed $\Omega_{\rm eq}/\Omega_{\rm c}$ do not change very much 
for $0<q<1$.  

This is one of the motivations for introducing new rotation laws 
(\ref{eq:rotlaw12}) whose Newtonian limits are the expressions on the
right in equation (\ref{eq:rotcurve012}).  For the case with the minus sign 
in Eq.~(\ref{eq:rotlaw12}), the slope of the rotation curves 
changes gradually from convex downward to upward as the value of the positive 
index $p$ decreases.  This is shown in panel (d) of 
Fig.~\ref{fig:rotcurve012} for the Newtonian limit.  
For the case with the plus sign in Eq.~(\ref{eq:rotlaw12}), 
the rotation curves monotonically increase along the equatorial 
radius, and their Newtonian limit is plotted in the panel (e) of 
Fig.~\ref{fig:rotcurve012}.  
As mentioned in the Introduction, these rotation laws with minus and 
plus signs may be used for modeling the evolution of the angular momentum 
distribution of the core of PNS and HMNS, or of neutron stars spinning up
because of fall back accretion.  

For the case of HMNS formed after BNS mergers, results of the 
simulations suggest that $j(\Omega)$ could become a multi-valued function.  
Therefore, we propose the integrability condition to be $\Omega$ as 
a function of $j$, $\Omega=\Omega(j)$, instead of $j=j(\Omega)$.
\footnote{
Eq.~(\ref{eq:Euler}) is also written 
$\frac{\na_\alpha p}{\epsilon+p}-\na_\alpha \ln u^t +j\na_\alpha\Omega=0$, 
and the relativistic von Zeipel's theorem states that 
the coincidence of the surfaces of constant energy density $\epsilon$ and 
pressure $p$ is guaranteed if and only if $f(j,\Omega)=0$,
where $f$ is a function of $\Omega$ and $j=u^t u_\phi$ only 
(or $f(l,\Omega)=0$ where $l:=-\frac{u_\phi}{u_t}=\frac{j}{1+j\Omega})$
\cite{1971AcA....21...81A}.
}
Accordingly the integral in Eq.~(\ref{eq:1stint}) should be rewritten,
\beq
\int jd\Omega = \int j \frac{d\Omega}{dj}dj.  
\label{eq:intOJ}
\eeq
Then, we propose two rotation laws, 
\beq
{\Omega(j;p,q,A,B)}
={\Omega_{\rm c}}\frac{1+\left(j/B^2\Omega_{\rm c}\right)^p}
{1+\left(j/A^2\Omega_{\rm c}\right)^{q+p}}.
\label{eq:rotlawOJ}
\eeq
\beq
{\Omega(j;p,A,B)}
={\Omega_{\rm c}}\left[1+\left(\frac{j}{B^2\Omega_{\rm c}}\right)^p\right]
\left(1-\frac{j}{A^2\Omega_{\rm c}}\right).
\label{eq:rotlawOJjco}
\eeq

In Fig.\ref{fig:rotcurveOJ}, rotation curves in the Newtonian 
limit are plotted for selected indices $(p,q)=(1,3)$ and $(2,3)$ 
for Eq.~(\ref{eq:rotlawOJ}), and $p=1$ and $1.5$ for 
Eq.~(\ref{eq:rotlawOJjco}).  In these curves, we determine the 
parameters $A$ and $B$ by setting the ratio of 
the maximum value of $\Omega$, $\Omega_{\rm max}$ to that 
at the rotation axis $\Omega_{\rm c}$, as well as the ratio 
$\Omega_{\rm eq}$ to $\Omega_{\rm c}$ to be a given constant.  
In Fig.\ref{fig:rotcurveOJ}, these ratios are set as 
$\Omega_{\rm max}/\Omega_{\rm c} = 2$ and 
$\Omega_{\rm eq}/\Omega_{\rm c} = 0.5$.  

In both (\ref{eq:rotlawOJ}) and (\ref{eq:rotlawOJjco}), 
the index $p$ controls the growth of rotation curves near the 
rotation axis.  This is analogous to the rotation law (\ref{eq:rotlaw12}) 
with the positive sign, but for (\ref{eq:rotlawOJ}) and (\ref{eq:rotlawOJjco}), 
$\Omega\sim \varpi^{2p}+\mbox{constant}$. 
The index $q$ in Eq.~(\ref{eq:rotlawOJ}) controls 
the asymptotic behavior of $\Omega(\varpi)$, in particular the index value 
$q=3$ results in the Kepler rotation law in the Newtonian limit.  
For non integer values of $(p,q)$, the hypergeometric 
function appears in the first integral.  Therefore, 
in actual applications of the rotation law (\ref{eq:rotlawOJ}), 
we choose indices $(p,q)=(1,3)$ and $(2,3)$ only.  
With these choices, the integral (\ref{eq:intOJ}) becomes 
a little lengthy but analytic expressions 
in terms of elementary functions exist.  

Rotation laws of the envelope or disk of matter ejected during 
events of core collapses or BNS mergers seem more likely to follow 
asymptotically the $j$-constant law as the remnants evolve toward 
axisymmetric configurations.  Model (\ref{eq:rotlawOJ}), 
however, can not reproduce $j$-constant rotation in a region 
$A \alt \varpi \leq R_0$.  
Such a $j$-constant law in the envelope can be achieved by 
rotation law (\ref{eq:rotlawOJjco}).  For Eq.~(\ref{eq:rotlawOJjco}), 
one can integrate Eq.(\ref{eq:intOJ}) analytically for 
any arbitrary values of $p$ in terms of elementary functions.  
When the constant $B$ is set to be $B>A$ in order to have 
$\Omega_{\rm max}\sim \Omega_{\rm c}$, 
this rotation law results in the $j$-constant law as in Eq.(\ref{eq:rotlawjc}).  
However, for Eq.(\ref{eq:rotlawOJjco}), $\Omega$ transits from 
uniform rotation to the $j$-constant rotation law more sharply 
than Eq.(\ref{eq:rotlawjc}).

These rotation laws (\ref{eq:rotlawjc}), (\ref{eq:rotlaw0}), 
(\ref{eq:rotlaw12}), (\ref{eq:rotlawOJ}), (\ref{eq:rotlawOJjco}) 
are incorporated successfully into our \cocal\ code 
\cite{cocal,Uryu:2016dqr}.  In the following calculations, 
we use the Isenberg Wilson Mathews formulation (thin sandwich 
formalism), which is based on the $3+1$ decomposition 
of the spacetime with the assumption of conformally flat spatial 
metric on the spacelike hypersurface.  We also assume 
that the high density matter of compact stars is a perfect fluid 
whose EOS is approximated by a polytropic EOS 
$p=K\rho^\Gamma$ with the index $\Gamma=2$ (see Table I).  
Further details on the formulation and numerical method can be 
found in our previous papers \cite{cocal,Uryu:2016dqr}.  
\begin{table}
\caption{The maximum mass and compactness of 
a spherically symmetric (Tolman-Oppenheimer-Volkov) solution 
for the case of the polytropic EOS $p=K\rho^\Gamma$ with $\Gamma=2$.  
$M_0$ is the rest mass, $M$ the gravitational mass, and 
$M/R$ the compactness ($R$ the circumferential radius).  
The constant $K$ is chosen so that $M_0=1.5$ at $M/R=0.2$.
To convert the value of $\rho_c$ in cgs units to that in $G=c=\Msol=1$ units, 
divide the value by 
$\Msol(G\Msol/c^2)^{-3}\approx 6.176393\times 10^{17} \mbox{g cm}^{-3}$.
}  
\label{tab:TOV_solutions}
\begin{tabular}{cccccc}
\hline
$\Gamma$ & $(p/\rho)_c$ & $\rho_c\ [\mbox{g/cm}^3]$ & $M_0$ & $M$ & $M/R$  \\
\hline
$2$ & $0.318244$ & $2.76957\times 10^{15}$ & $1.51524$ & $1.37931$ & $0.214440$  \\
\hline
\end{tabular}
\end{table}

\begin{table}
\caption{Parameters of the rotation laws used for computing equilibrium solutions 
presented in Fig.\ref{fig:solutions}.  Models I to V and DR correspond to the pair 
of panels in Fig.\ref{fig:solutions} in order from top to bottom in the 
left column, then top to bottom in the right column.  
Model DR is one of the HMNS solutions calculated in BSS \cite{Baumgarte:1999cq}.  
}
\label{tab:rotlaw_param}
\begin{tabular}{ccccccc}
\hline
Model & Rotation law & $p$ & $q$ & $A/R_0$ & 
$\Omega_{\rm eq}/\Omega_{\rm c}$ & $\Omega_{\rm max}/\Omega_{\rm c}$  \\
\hline
I   & (\ref{eq:rotlaw12}) (minus)     & $1/4$ & --- & --- & $0.5$ & --- \\
II  & (\ref{eq:rotlaw12}) (plus)\ \ \ & $1/2$ & --- & --- & $0.5$ & --- \\
III & (\ref{eq:rotlaw0})              & ---   & $4$ & $10^{-3/2}$ & --- & ---   \\
IV  & (\ref{eq:rotlawOJ})             & $1$   & $3$ & --- & $0.5$ & $2$ \\
V   & (\ref{eq:rotlawOJjco})          & $3/2$ & --- & --- & $0.5$ & $2$ \\
DR & (\ref{eq:rotlaw0})               & ---   & $1$ & $1$ & ---   & --- \\
\hline
\end{tabular}
\end{table}

\begin{figure*}
\begin{center}
\includegraphics[height=40mm]{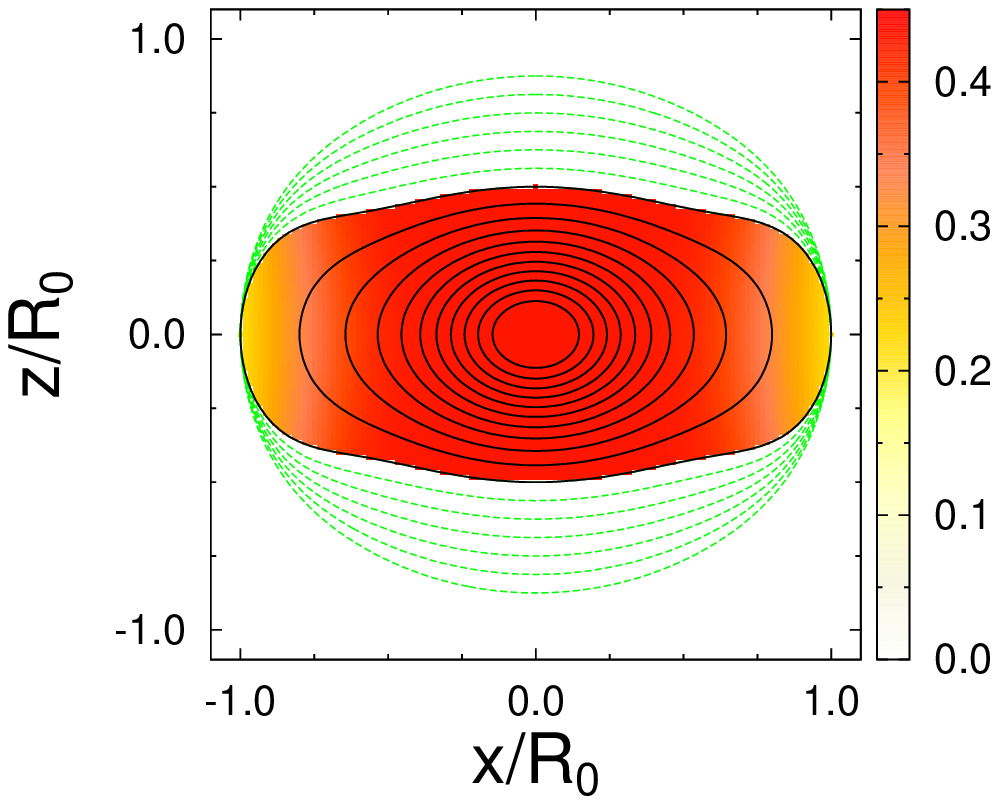}\ \ 
\includegraphics[height=40mm]{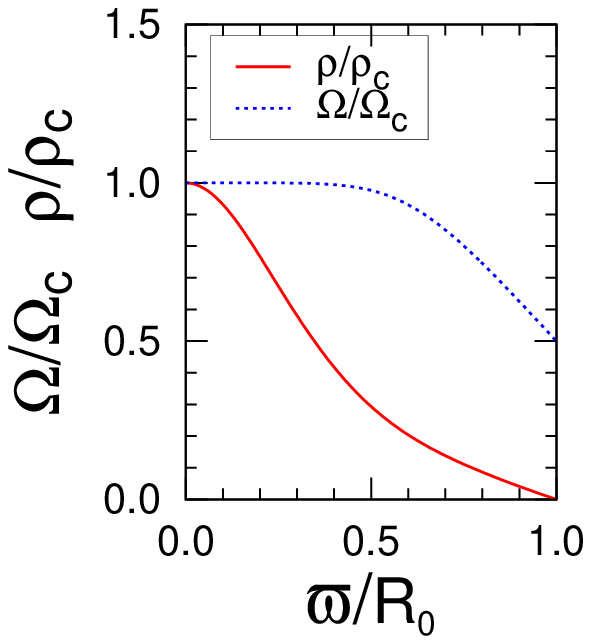}\ \ 
\includegraphics[height=40mm]{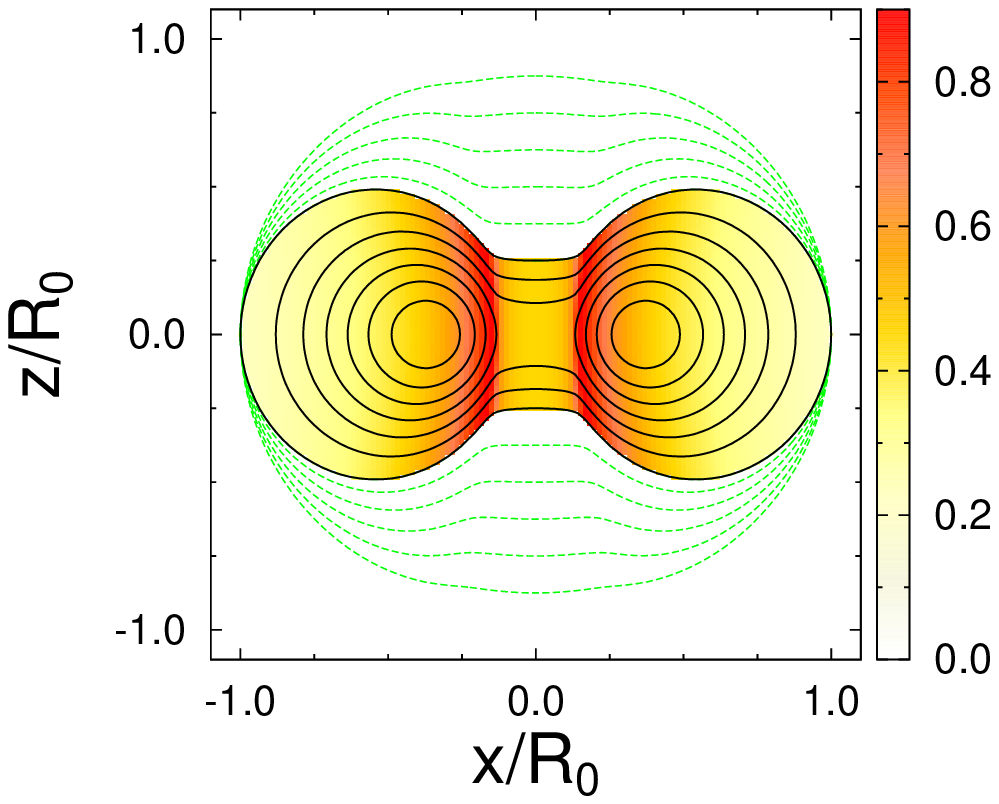}\ \ 
\includegraphics[height=40mm]{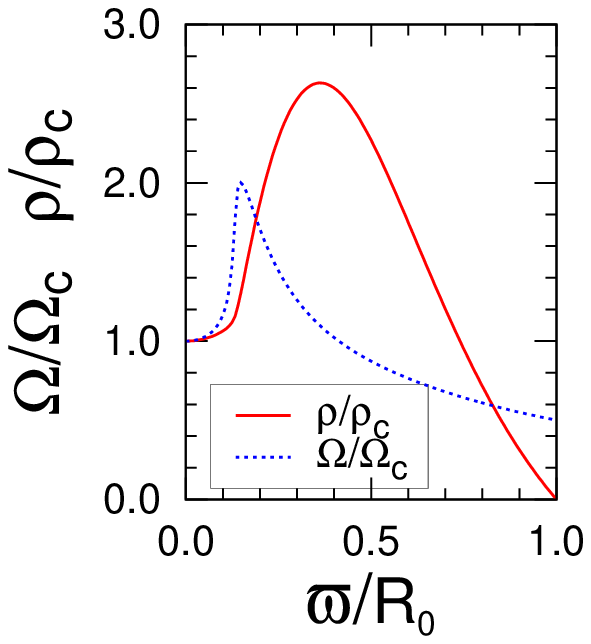}
\\
\includegraphics[height=40mm]{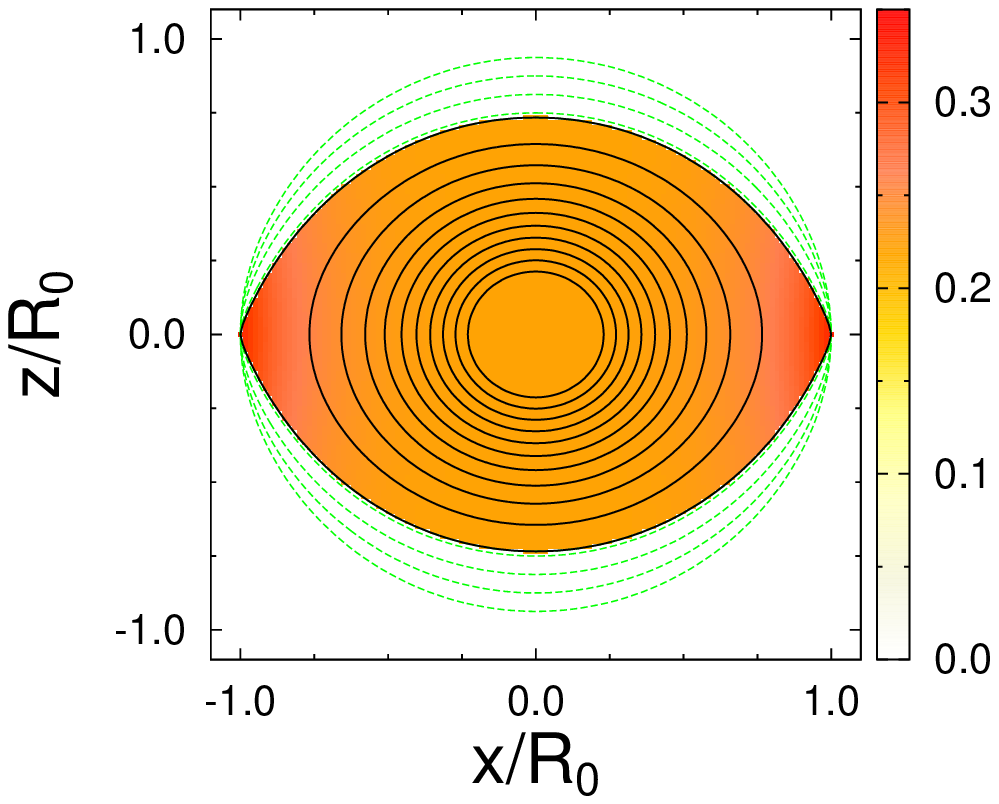}\ \ 
\includegraphics[height=40mm]{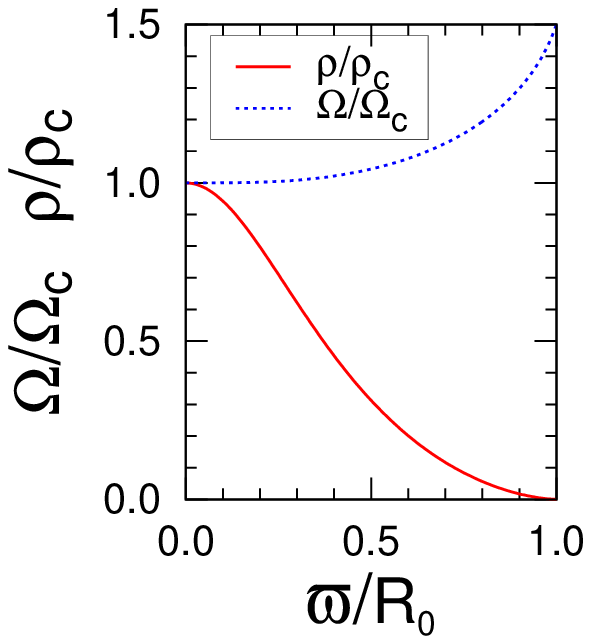}\ \ 
\includegraphics[height=40mm]{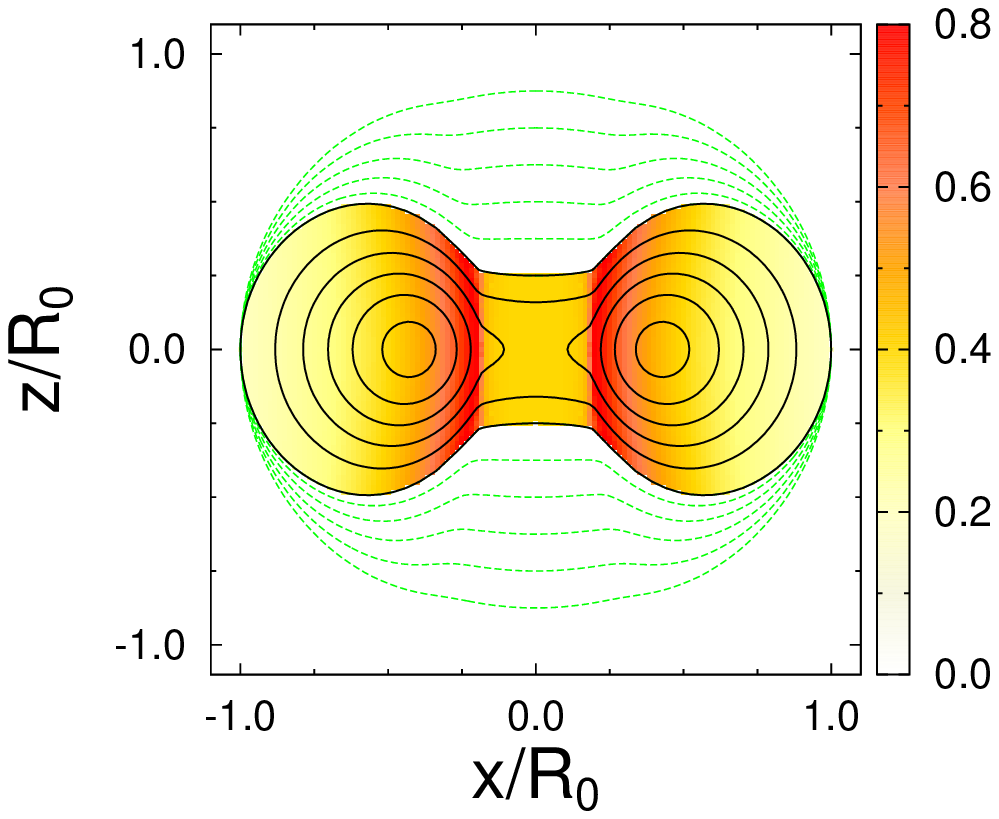}\ \ 
\includegraphics[height=40mm]{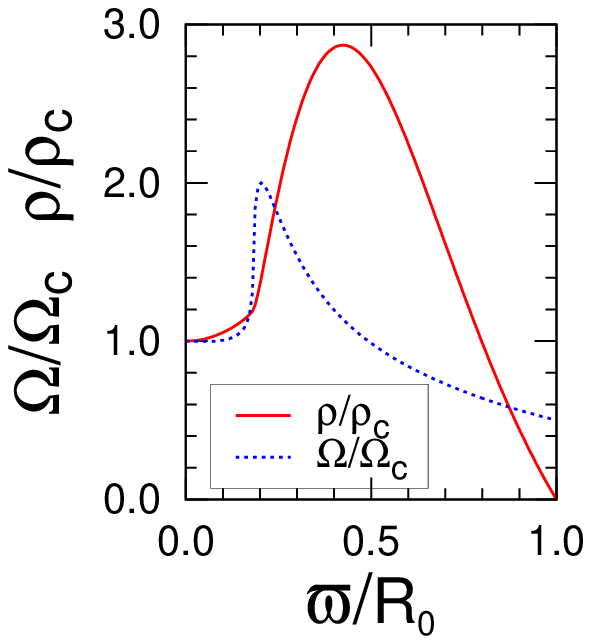}
\\
\includegraphics[height=40mm]{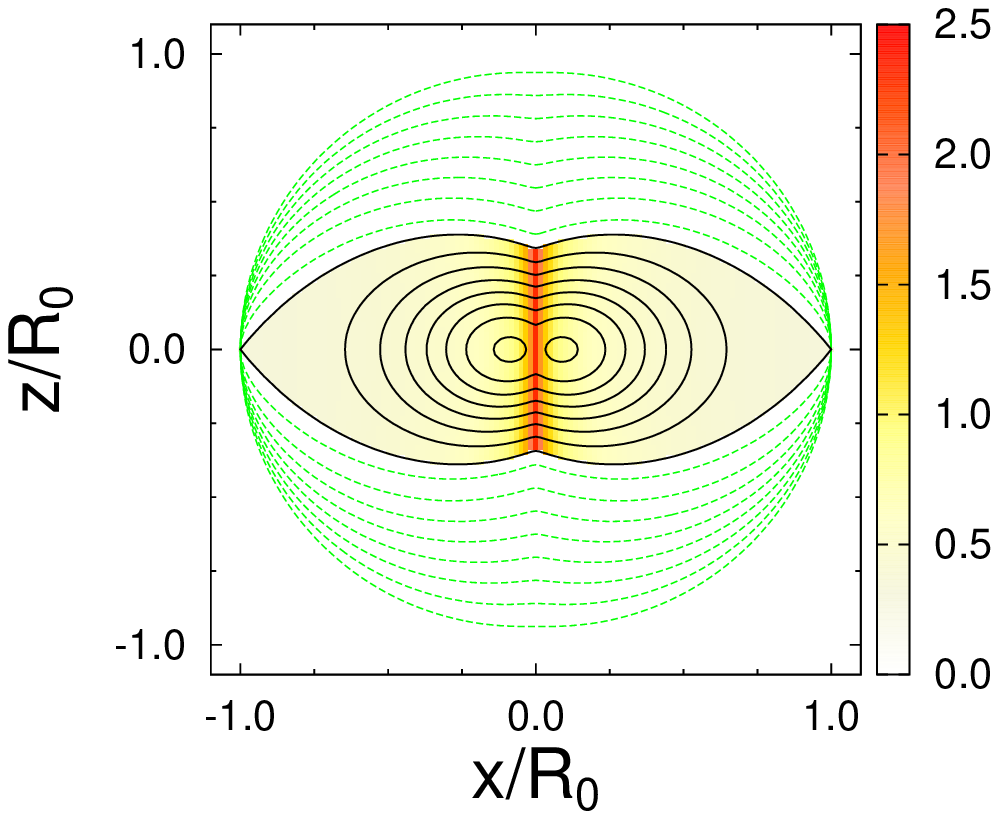}\ \ 
\includegraphics[height=40mm]{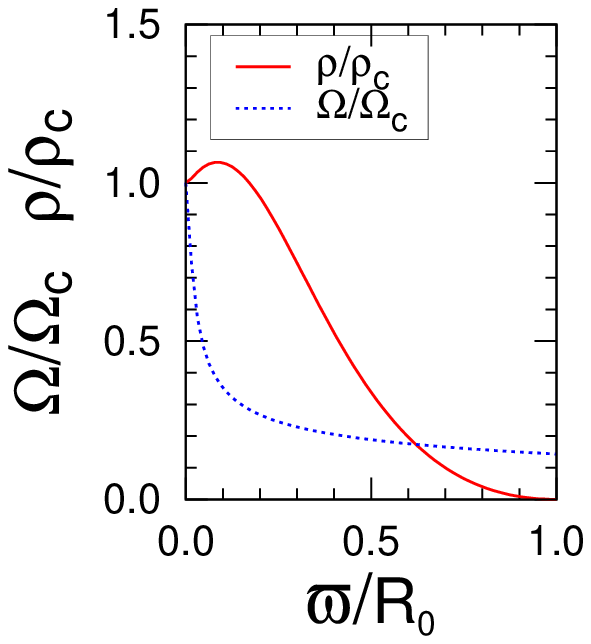}\ \ 
\includegraphics[height=40mm]{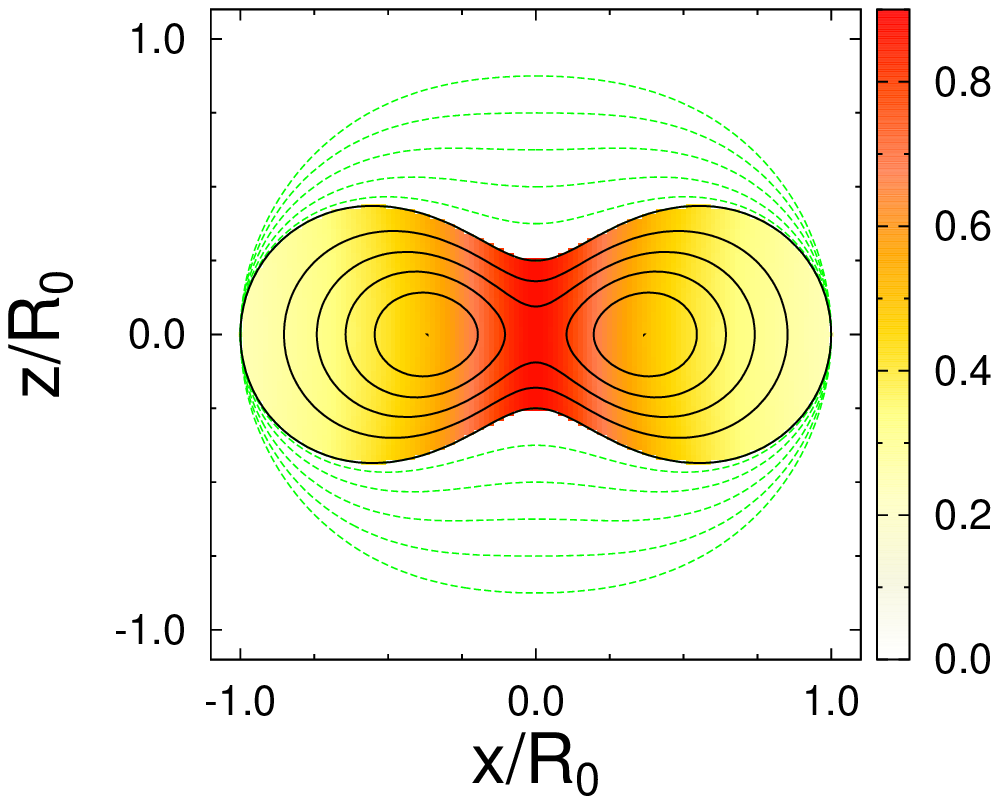}\ \ 
\includegraphics[height=40mm]{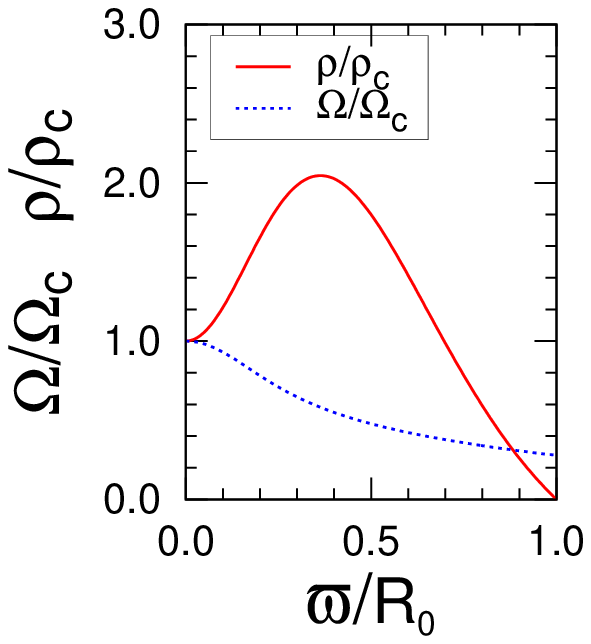}
\caption{
Equilibriums of differentially rotating compact stars.  
The set of model parameters for the rotation law for each pair of panels is 
described in Table \ref{tab:rotlaw_param}.  The pairs of panels 
from top to bottom in the left column, then from top to bottom 
in the right column correspond to Models I to V and DR in 
Table \ref{tab:rotlaw_param}, respectively.  
The left panel of each pair shows the contours of the rest mass density 
$\rho$ (black solid curves), the color map for the angular velocity $\Omega R_0$ 
and the deformation sequence of the surface of the rotating star 
$R_s(\theta,\phi)$ (green dashed curves)
on the $xz$ (meridional) plane.  
In the right panels, plotted are the normalized rest mass density profiles 
$\rho/\rho_c$ and the normalized rotation curves $\Omega/\Omega_{\rm c}$ 
along the equatorial axis ($x=\varpi \sin\theta$ with $\theta =\pi/2$).  
}
\label{fig:solutions}
\end{center}
\end{figure*}
\begin{figure}
\begin{center}
\includegraphics[height=55mm]{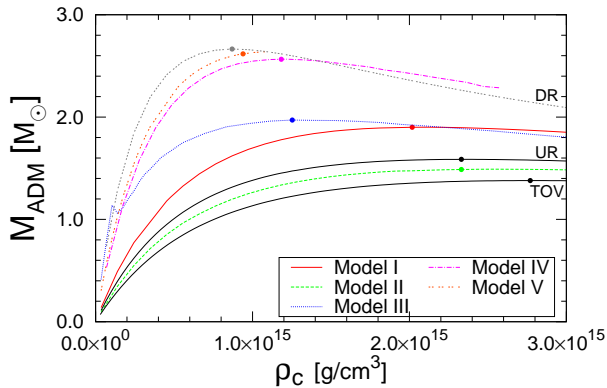}
\caption{
$\Madm$ is plotted with respect to the maximum rest mass density 
$\rho_c$ for differentially rotating Models 
I to V in Table \ref{tab:rotlaw_param}.  For reference, 
Model DR in Table \ref{tab:rotlaw_param} \cite{Baumgarte:1999cq}, 
the uniformly rotating Model UR and the spherically 
symmetric (TOV) model are also plotted.  
The curves for Models II, III and UR correspond to their 
maximum deformation sequences.  Those of Models IV, V, DR 
correspond to sequences with a fixed axis ratio 
$R_z/R_0=0.25$, and Model I with $R_z/R_0=0.5$.  
Each solution in Table \ref{tab:DiffRot_solutions} 
and the maximum mass of the TOV solution Table \ref{tab:TOV_solutions} 
is marked by a circle on each curve.  
}
\label{fig:maxmass}
\end{center}
\end{figure}

\begin{table*}
\caption{Selected solutions for various rotation law models.  
Models I--V and DR correspond to the rotation laws listed in Table \ref{tab:rotlaw_param}.  
The uniformly rotating Model UR is also listed for reference.  
Listed quantities are the equatorial and polar radii in proper length $\bar{R}_0$ 
and $\bar{R}_z$, the maximum density $\rho_c$, the angular velocity near the rotation 
axis $\Omega_{\rm c}$, the ADM mass $\Madm$, the rest mass $M_0$, 
the proper mass $M_{\rm P}$, the angular momentum $J$, the ratio of the kinetic 
to gravitational energy $T/|W|$, the virial constant $I_{\rm vir}$, and the Komar mass $\MK$.  
Details of the definitions are found in \cite{cocal}.  
}
\label{tab:DiffRot_solutions}
\begin{tabular}{cccccccccccc}
\hline
Model &  
$\bar{R}_0$ &  $\bar{R}_z/\bar{R}_0$ &  $\rho_c\ [\mbox{g/cm}^3]$ 
& $\Omega_{\rm c}$ & $\Madm$ & $M_0$ & 
$M_{\rm P}$ & $J/\Madm^2$ & $T/|W|$ & 
$I_{\rm vir}/|{\cal W}|$ & $|1-\MK/\Madm|$ 
\\
\hline
I  & 
12.100   & 0.54465 & $2.0178\times 10^{15}$ & 
0.056201 & 1.90066 & 2.09398 & 
2.28918  & 0.79695 & 0.1623 & 
$7.13\times 10^{-4}$ & $5.14\times 10^{-5}$ \\
II &  
9.7507  & 0.76609 & $2.3309\times 10^{15}$ & 
0.034174 & 1.48792 & 1.63360 & 
1.82150 & 0.43227 & 0.0494 & 
$8.46\times 10^{-4}$ & $1.59\times 10^{-5}$ \\
III& 
14.921  & 0.38561 & $1.2531\times 10^{15}$ & 
0.226992 & 1.97116 & 2.16320 & 
2.31631 & 0.84818 & 0.1854 & 
$1.06\times 10^{-3}$ & $4.80\times 10^{-5}$ \\
IV & 
12.523  & 0.26763 & $1.1830\times 10^{15}$ & 
0.062582 & 2.56462 & 2.84625 & 
3.07421 & 0.88534 & 0.2464 & 
$1.60\times 10^{-4}$ & $2.67\times 10^{-4}$ \\
V  & 
13.501  & 0.26010 & $8.6983\times 10^{14}$ & 
0.046783 & 2.59667 & 2.86195 & 
3.03360 & 0.91444 & 0.2461 & 
$8.59\times 10^{-5}$ & $2.00\times 10^{-4}$ \\
DR & 
14.230  & 0.26688 & $8.6985\times 10^{14}$ & 
0.100879 & 2.66479 & 2.94986 & 
3.12656 & 0.95184 & 0.2589 & 
$1.44\times 10^{-4}$ & $2.09\times 10^{-4}$ \\
UR & 
11.141 & 0.63972 & $2.3309\times 10^{15}$ & 
0.044776 & 1.58665 & 1.74325 & 
1.93925 & 0.56764 &  0.0832 & 
$8.87\times 10^{-4}$ & $2.54\times 10^{-5}$ \\
\hline
\end{tabular}
\end{table*}
\section{Results}
\label{sec:results}

In Fig.\ref{fig:solutions}, solutions of representative 
models for compact stars with various differential rotation 
laws are presented.  In the left panel of each pair in the figure, 
shown are the contours for the rest mass density $\rho$, 
the color map for the distribution of the angular velocity 
$\Omega$ (normalized to the equatorial radius $R_0$), 
and the deformation sequence of the stellar surfaces 
(dashed green curves) in the $xz$ (meridional) plane
\footnote{
The degree of deformation of 
an axisymmetric star may be measured by 
the ratio of the axes along the polar and equatorial radius $R_z/R_0$.  
Each dashed green curve in each panel of Fig.\ref{fig:solutions} 
corresponds to the shape of a star in the $xz$ plane with a fixed $R_z/R_0$.  
(Each dashed green curve corresponds to the surface of a different solution, 
and sequences of dashed green curves are deformation sequences of each 
differentially rotating model.)  The contours for $\rho$ 
and the color map for $\Omega$, as well as the corresponding right panels, 
are those of the solution with the largest deformation among those computed 
for each model.  
}.  
In the right panels, the rest mass density profile 
$\rho/\rho_c$ and the rotation curve $\Omega/\Omega_{\rm c}$ 
normalized to the central value of each quantity are plotted along 
the radial coordinate $\varpi$ in the equatorial plane.  

Selected parameters 
for the differential rotation laws corresponding to the models 
in Fig.\ref{fig:solutions} are summarized in Table\ref{tab:rotlaw_param}.  
Physical quantities of these selected solutions are listed in Table 
\ref{tab:DiffRot_solutions}.  Also listed are, for reference, 
the differentially and uniformly rotating models DR and UR, respectively, 
calculated for the same EOS.  Model DR is one of the HMNS models used in 
BSS \cite{Baumgarte:1999cq}.  
The solutions of Models I, II, III, and UR in Table 
\ref{tab:DiffRot_solutions} are those closest to the maximum 
mass model of each rotation parameter set.  
Those of Models IV--V and DR are close to 
the maximum mass model among the calculated equilibriums.  
Solution sequences of these models are expected to continue to 
larger deformations and eventually to completely toroidal configurations.  
Such toroidal equilibriums may support a much larger mass than 
the presented solutions.  
Also note that, in Models II and III, the rotation 
of the fluid near the equatorial surface is close to 
the Kepler limit.  

In Fig.\ref{fig:maxmass}, the Arnowitt-Deser-Misner (ADM) mass $\Madm$ 
is plotted with respect to the maximum of the rest mass density $\rho_c$.  
The curves for Models II, III, and UR are the extrapolated maximum values 
of $\Madm$ for each value of $\rho_c$.  For Models IV, V, and DR 
which exhibit toroidal configurations, the curves are those of fixed 
axis ratio $R_z/R_0=0.25$, and for Model I, the curve is for $R_z/R_0=0.5$.  
Each solution in Table \ref{tab:DiffRot_solutions} is marked by 
a circle in the plots.  
It can be seen clearly that hypermassive solutions exist for 
differential rotation Models I, III--V, and DR, while only 
supramassive solutions exist for Model II.  \footnote{We call 
a compact star solution hypermassive if its 
mass is larger than that of the maximum mass of the uniformly rotating 
solution, and supramassive if its solution is smaller than the maximum 
mass of the uniformly rotating solution, but larger than the maximum mass
of the spherically 
symmetric static solution \cite{Baumgarte:1999cq}.}

In Model I (top left column of Fig.\ref{fig:solutions}), 
the differential rotation law is nearly constant in the core,
and decreases to $50 \%$ of the core value 
at the equatorial surface.  This type of rotation law 
may be realized in the PNS formed 
as a result of core collapse.  Because of the 
uniform rotation of the core, the deformation of equilibrium 
solutions appear to be limited: examining the density contours 
of several solutions slightly more deformed than that of Model I 
in Fig.\ref{fig:solutions}, we found that, 
although the surface of 
the star is not close to the Kepler limit, the largely 
deformed inner part reaches break-up because of rapid 
rotation, and hence the deformation sequence of 
spheroidal equilibriums is terminated.

For Model II (the second panel from the top in 
the left column in Fig.\ref{fig:solutions}), 
the equatorial angular velocity $\Omega_{\rm eq}$ of the star is 
$50\%$ higher than the central value $\Omega_{\rm c}$.  
In such a case, the maximum deformation 
can not be larger than that of the uniformly rotating Model UR.  
Although Model II is less deformed than Model UR, 
$\Omega_{\rm eq}$($=2\Omega_{\rm c}$ in this case) 
at the maximally rotating model (Kepler limit) is about $53\%$ 
{\it larger} than $\Omega$ of the uniformly rotating model (in normalized 
angular velocity $\Omega \Madm$ is $43\%$ larger), 
while the $T/|W|$ of Model II is $59\%$ {\it smaller} than that 
of Model UR.  Such solution may become 
a model for neutron stars spinning up because of equatorial 
accretion.  In more realistic situations, it might be necessary 
to apply external pressure at the surface of the star, which 
may allow more rapid rotation than the presented model.  

Model III (the third panel from the top of the left column 
in Fig.\ref{fig:solutions}) is an example for which 
the rotation curve is a power of $\varpi$.  
In this model, we choose $q=4$ in Eq.~(\ref{eq:rotlaw12}), which becomes 
$\Omega \sim \varpi^{-2/q} = \varpi^{-0.5}$ in the Newtonian limit.  
We have confirmed that this power law is reproduced in the less massive 
solutions, but in the presented solution which is close to 
the maximum mass, the index appear to be $\Omega \sim \varpi^{-0.4}$, 
which is considered to be a relativistic effect in isotropic 
coordinates.  

In the pairs of panels in the right column of Fig.\ref{fig:solutions}, 
equilibrium models largely deformed to become toroidal density 
distributions are presented.  Model IV (top panels) is the result for 
Eq.(\ref{eq:rotlawOJ}) with $(p,q)=(1,3)$, and Model V 
(middle panels) for Eq.(\ref{eq:rotlawOJjco}) with $p=3/2$.  
Model DR (bottom panels) 
is the model with Eq.(\ref{eq:rotlawOJ}) with $q=1$, which is the 
same as rotation law (\ref{eq:rotlawjc}), and is shown for 
comparison.  Model DR has been commonly used for studying HMNS 
and other relativistic differentially rotating stars including PNS 
\cite{RNSdiff,Baumgarte:1999cq,Morrison:2004fp,Kaplan:2013wra}.  
Solution sequences of increasing deformation of these three models 
seem not to terminate at a certain axis ratio, but continue to 
toroidal solutions.  This allows masses as high as, or even higher than, 
twice the maximum mass of the spheroidal solution.  

Since these differential rotation profiles allow toroidal distributions of 
mass, these three solutions are qualitatively similar.  
However, the detailed structure of the solutions, especially near 
the rotation axis, depends on the rotation curves.  
The rotation curve is flatter for larger index $p$, 
and accordingly, the distance between the rotation axis and 
the maximum density (center of the toroidal density distribution) 
becomes wider.  

The overall features of the rotation curve of each compact star presented in 
Fig.~\ref{fig:solutions} are qualitatively the same as the corresponding 
Newtonian rotation curves shown in Figs.~ \ref{fig:rotcurve012} 
and \ref{fig:rotcurveOJ}.  Differences in the details of the profiles may 
be due to the use of isotropic coordinates on the spacelike hypersurface, 
for which the coordinate length becomes relatively short in 
the region of stronger gravity (with higher density).

%
%
\section{Discussion}
\label{sec:discussion}

In recent papers \cite{SKS,Radice:2017zta}, 
fully numerical relativity simulations 
for high density matter associated with large 
shear viscosity have been performed.  
In \cite{SKS}, the evolution of rotation 
profiles due to viscosity was shown for the HMNS formed after 
BNS mergers.  
Because of the assumption of high viscosity, the rotation 
profile evolves towards uniform rotation in a short 
timescale, about $\sim 20$msec, and hence the rotating 
HMNS stably evolves in such short timescale.  However, 
if the viscosity is not strong enough, the evolution 
timescale of such HMNS may be longer, and then 
the stability of each rotating model with a certain 
rotation profile needs to be examined.  Our equilibrium 
models will be useful for studying the stability 
of such HMNS.  
As a step for such a study, one can investigate 
the bar mode instability of differentially rotating 
neutron stars with the proposed new rotation laws, 
as it has been done limitedly to rotation law (\ref{eq:rotlawjc}) 
in \cite{barmode}.

We have proposed new function forms of integrability 
conditions for differential rotation laws of relativistic 
compact stars in equilibrium.  It is possible to develop 
further variations of function forms for other 
rotation laws.  More practically, one could prepare 
interpolating or fitting functions for a data table 
of $j(\Omega)$ or $\Omega(j)$ obtained from the results 
of numerical simulations.  
In this paper, we have concentrated on the effect of 
differential rotation on the increase of the maximum mass 
seen in the HMNS.  
However, in more realistic situations, the increase of the mass 
of HMNS is due also to the thermal part of the EOS.  
It is therefore important to include the thermal part of the EOS 
to develop more realistic equilibrium models of such 
PNS or HMNS for distinguishing the contributions from the 
thermal pressure and the differential rotation to 
the mass excess.  
As for GW170817, a search for a rotating neutron star remnant has been 
performed in the data following GW170817 and the upper limits of 
gravitational wave signal is obtained which is of an order of magnitude 
larger than expected.  The signal would be detected by LIGO/VIRGO of 
the design sensitivity and the such future detectors as KAGRA 
\cite{Abbott:2017eaw}.  
Such extensions of the present method and their applications to 
actual compact objects are the next step of our future works.  

\acknowledgments
This work was supported by 
JSPS Grant-in-Aid for Scientific Research(C) 15K05085, 
17K05447, 26400267, 26400274, NSF Grants PHY-1602536, PHY-1662211,
NASA Grants NNX13AH44G, 80NSSC17K0070.  
%

%

\end{document}